\documentclass[referee]{aa}
\usepackage{graphicx}
\usepackage{natbib}
\bibpunct{(}{)}{;}{a}{}{,}
\begin{document}

   \title{Transient pulsar dynamics in hard x-rays: 
          Prognoz 9 and GRIF "Mir" space experiments data}

   \author{M.I. Kudryavtsev
          \inst{1}
          \and
          S.I. Svertilov\inst{2}
         \and V.V. Bogomolov\inst{2}
          }

   \offprints{S.I. Svertilov}

   \institute{Space Research Institute, Russian Academy of Science, 
              Profsoyuznaya st. 84/32, Moscow 117810, Russia
             \email{sis@coronas.ru}
         \and
             Skobeltsyn Institute of Nuclear Physics, 
             Moscow State University, Vorob'evy Gory, Moscow 119899, Russia
             \email{sis@coronas.ru}
             }

   \date{ }

\titlerunning{Transient pulsar dynamics in hard x-rays}
\authorrunning{M.I. Kudryavtsev, S.I. Svertilov, V.V. Bogomolov}

\abstract{
The long-term observations of the Galactic Centre as well as the Galactic 
anti-Centre regions in hard X-rays (10-300 keV) were made in experiments 
on board Prognoz-9 satellite and "Mir" orbital station (GRIF experiment). 
Some transient pulsars including A0535+262, GS1722-36, 4U1145-619, A1118-615, 
EXO2030+37, Sct X-1, SAX J2103.5+4545, IGR 16320-4751, IGR 16465-4507 were 
observed. The pulsation flux components of A0535+26 and GS1722-36 X-ray 
emission were revealed at significant level. For other observed pulsars 
the upper limits of pulsation intensity were obtained. The mean pulsation 
profiles of A0535+26 in different energy ranges as well as the energy spectra 
were obtained at different stages of outburst decreasing. The pulsation 
intensity-period behavior does not contradict the well-known correlation 
between spin-up rate and X-ray flux, while the stable character of the energy 
spectrum power index indicates on the absence of thermal component. The energy 
spectrum and mean pulsation profiles were also obtained for one time interval 
of GS1722-36 observations. The upper limits of pulsation fluxes obtained for 
other observed transient pulsars at the orbital phases more than 0.14 
correspond the quiescent state or final stage of the first type outburst. }


   \maketitle
%

\section{Introduction}
	It is well known that so called X-ray pulsars are associated with strongly magnetized rotating 
neutron stars (NS), which are characterized by great energy release as the result of accretion 
(accretion-powered pulsars) or rotation (rotation powered pulsars). The X-ray accretion-powered 
pulsars are generally divided into different classes based on the spectral type classification of the 
mass donor companion star \citep{Mer01}. The temporal behavior of such objects was always a 
problem of great interest for high-energy astrophysicists. Modern observations are usually made in a 
wide range of wavelengths from radio waves to gamma rays. While optical observations are in some 
respects more accurate, the observation of X-ray pulsars has provided an important information on 
the physics of NS, including the processes in the region of great energy release, and on the evolution 
of stars in binary systems. 

The temporal behavior of NS binary system in hard X-rays reflects the physical processes in 
matter and fields surrounding NS. The properties of such matter and field determine quasi-periodic 
variations, while the rotation of NS itself and of binary system components produce pulsation, 
orbital and super-orbital periodic changes in X-ray source luminosity. X-ray pulsations with pulse 
period of about dozens and hundreds seconds were clearly detected. Long-term variability of most 
of transient pulsars appears as their flaring activity. 

The most of transient 
X-ray pulsars have Be (or Oe) star companions. Accreting NSs in Be systems typically have long 
periods and eccentric orbits. It is to be thought, that the source of accreting material is the slow, 
dense stellar wind, presumably from the circumstellar disk confined to the equatorial plane of the 
rapidly rotating Be star. The Be-transients display a marked correlation between their spin and 
orbital periods. This correlation arises from the fact that given identical companion masses and 
mass loss rates, NSs in binary systems with lower orbital periods are further away from their 
companions, these leading to lower mass accretion rates and higher equilibrium periods \citep{Bil97}. 
Various instabilities in the accretion process lead to the flaring activity of some kind of X-
ray pulsars. It is known that X-ray behavior of Be transient pulsars is characterized by regular 
increases of the X-ray flux modulated with the orbital period. Such recurrent flares (type 1 
outbursts) with typical times about days often begin soon after periastron. The typical X-ray 
luminosity of type 1 outbursts in BeX systems is about $10^{36}$ erg/s \citep{Bla04}. Besides 
the series of periodically occurring outbursts, the so-called giant outbursts with higher luminosity
(type 2 outbursts) were observed. The type 1 outbursts are associated 
with the direct wind accretion of the Be circumstellar disk, while the type 2 outbursts can be 
described by the disk-fed accretion after the bound material has collapsed into a standard but 
temporary accretion disc \citep{Ste86,Rag98}. 
The outburst peaks occur at phase 0-0.5, depending on the wind characteristics and orbital 
eccentricity. The bright giant outbursts have high spin-up rates, longer duration, and often 
have peak at an orbital phase delayed relative to the mean normal outburst X-ray maximum (e.g. 
4U0115+63 \citep{Whi89}, 3A0535+26, \citep{Bil97}). 

Extensive data on periodicity in hard X-rays from many transient pulsars were obtained 
during the BATSE experiment onboard the Compton orbital observatory (CGRO) \citep{Bil97}. 
As the result of BATSE CGRO experiment no association had been observed 
between giant and normal outbursts. It was found that many of the giant outbursts are in the middle, 
or followed by, series of normal outbursts. Sometimes type 2 outbursts last for several orbital cycles 
(e.g. V0332+53, \citep{Ste86}). The phasing of this outbursts should be dictated 
mainly by the time variability of the Be star mass outflow rate. It is still unclear what causes 
the giant outbursts. They may be observed if the circumstellar disk undergoes a sufficiently large increase 
in its radial extent and density to intersect the NS orbital path. If a disk can be sustained between 
giant outbursts, it may be also present during normal outburst. In this case the large tidal torque 
experienced by the disk during periastron passage could explain the repeating normal outburst 
\citep{Bil97}. Not far ago it was proposed that the thermal disk instability like one causing 
dwarf nova outbursts affect accretion disk around Be X-ray pulsars and could be the cause of the 
giant outburst \citep{VanPar96,Kin96}.

The pattern of X-ray outbursts is affected by the size, eccentricity and orientation of the NS's 
orbit with respect to the Be star. The orbit could be complanar with the Be star circumstellar disk or 
offset such a way that the NS may pass through the disk. In some case the consequent outbursts can have 
different intensities (e.g. XTE J1946+274, \citep{Cam99} ). This might suggest that the NS is 
orbiting the Be companion in an inclined orbit crossing twice the Be disk plane and giving rise to 
two outbursts per orbit. Two flares per orbit also observed in 4U1907+097 \citep{Mak84}
and in GRO J2058+42 \citep{Wil98}.
Be transient pulsars also display long-term X-ray variability with typical times about 
months, consisting of low and high-activity X-ray states. By this the type 1 outbursts are only seen 
during bright states \citep{Bay02}.

Nevertheless moderate and pulsating X-ray flux was detected at various orbital phases of 
Be transient binary systems, from A0535+26, for example \citep{Mot91,Stee98}. 
In the RXTE-PCA observations this source was detected at a much 
lower luminosity of $(2.0-4.5)\cdot10^{33}$ erg/s at which so-called propeller effect is expected. It 
means that accretion onto the NS surface is inhibited by the centrifugal action of the rotating 
magnetosphere. Thus the X-ray luminosity that was still detected in this regime can be ascribed to 
material leaking through the magnetosphere or thermal emission from the heated core of the NS 
\citep{Neg00}. The BeppoSax-NFI observations of A0535+262 in quiescence show that 
X-ray pulsations are still present for luminosity as low as $2.0\cdot10^{33}$ erg/s, but they have not been 
detected at lower luminosity $\sim1.5\cdot10^{33}$ erg/s while the nonzero moderate flux was measured 
\citep{Muk05}. This means at such luminosity a transition between centrifugal 
inhibition and direct accretion when a fraction of the disk material is going onto the surface of the 
NS along the magnetic field lines. The luminosity obtained in the BeppoSax-NFI observations gives 
for A0535+262 in quiescence an accretion rate $\dot{M} = 2.05\cdot10^{13}$ g/s and magnetospheric 
radius $r_m$ about $9.35\cdot10^9$ cm \citep{Muk05}.
 
In the case of type 1 outbursts caused by the wind-fed accretion, there will be less efficient 
angular momentum transfer compared to a standard transfer disk accretion \citep{Ina04}. On 
the contrary, the disk-fed accretion can transfer angular momentum either at the vicinity of the 
magnetosphere radius where the disc is disrupted and the material is channeled from the inner 
edge of the disc to the magnetic poles, or from the overall interaction of the accretion disc and 
magnetic field lines of the NS. Thus, the material torque from a prograde accretion disk, being 
proportional to mass accretion rate, always acts to spin-up the NS, while the contribution of 
magnetic torque from the magnetic field lines threading the disk outside the corotation radius is 
negative. The resultant torque can either be a spin-up or spin-down. The large and steady spin-up 
rates were seen during the giant outbursts. This implies the correlation between torque and observed 
X-ray flux, which is difficult to explain with direct wind accretion. Assuming observed X-ray 
luminosity is proportional to the bolometric luminosity, i.e. mass accretion rate, spin-up rate and 
X-ray flux correlation can be explained by accretion from accretion disk when the net torque is 
positive and of the order of the material torque \citep{Gho79,Gho93}. Thus, 
it means that transient accretion disks are forming during the giant outbursts \citep{Kri83}
\citep{Ste86,Mot91}. 

When the NS in a Be/NS binary system leaves the dense equatorial disc of the companion, 
the accretion disk can no longer be fed by the surrounding material. In this case, accretion disc may 
disappear an the NS may either continue to accrete from the non-equatorial wind of the companion 
or may enter the propeller phase \citep{Ill75}. In case accretion is the 
result of the companion's non-equatorial wind, it is possible to see erratic spin-up and spin-down 
episodes \citep{Bil97,Ina00}. In case wind of the companion 
does not cause accretion, propeller phase may set in, when spin pulsation cessation is accompanying 
by the flux decreasing \citep{Cui97}. However, the pulsations may not cease completely even 
in the propeller phase \citep{Neg00}. Spectral hardening, which was observed in the low 
state spectra of NS soft X-ray transients like Aql X-1 may be interpreted as the sign of propeller 
stage \citep{Zha98}. While anti-correlation of spectral power law index with the X-
ray flux (i.e. softening of the spectra) might be the sequence of mass accretion rate changes  
\citep{Mes83,Har84}. In this case, neither a transition to propeller stage nor 
an accretion change is needed to explain the softening in the spectrum with decreasing flux 
\citep{Ina04}. Consequently, decrease in mass accretion rate with a softening in the spectrum does 
not lead to any significant changes in pulse profiles and pulse fraction \citep{Bay02}. 
In the case of 2S 1417-62 spectral softening accompany the changes in pulse profiles and pulse 
fractions that may be the result of not only accretion rate changes but also accretion geometry changes 
for the low flux parts, which occur just before the periastron for which the NS should have accreted 
almost all of the accretion disc material around it.
Correlation between spin-up rate and X-ray flux in different energy bands has been observed 
in outbursts of different transient X-ray pulsar systems: EXO 2030+375 \citep{Wil02},
A0535+26 \citep{Bil97}, 2S 1845-024 \citep{Fin99}, GRO J1744-28 \citep{Bil97}, 
GRO J1750-27 \citep{Sco97}, XTE J1543-568 \citep{InZ01}, SAX J2103.5+4545 \citep{Bay02},
2S 1416-62 \citep{Ina04}. 
In the last source the correlation between spin-up rate and X-ray flux was found for both main 
outbursts and the following mini outbursts \citep{Ina04}. However, the correlation between 
torque and flux, which was observed in BATSE experiment for some Be X-ray pulsars \citep{Fin96a,Fin96b},
does not confirm the pulsar spin-up rate $\dot{\nu}$ 
dependence on accretion rate $\dot{M}$, such as $\dot{\nu} \sim \dot{M}^{6/7}$, which was 
predicted by accretion torque theory. It is still unclear if this disagreement can be explained by 
bolometric or beaming corrections \citep{Bil97}. The BATSE instrument does not measure 
bolometric flux, but only pulsed flux in definite bands, while the large changes in beaming fraction 
imply the changing pulse profiles. Thus, it seems clear that new data on transient pulsars dynamics 
implying the flux, pulse profile and rotation period measurements on the long-term database 
including different luminosity regimes are quite useful for further progress on these phenomenon 
understanding. 
Below we will discuss the results of the search and study of periodic processes in hard 
X-rays from some transient pulsars during the observations which were carried out on the Prognoz-9 
and "Mir" orbital station (OS) missions.

\section{Monitor observations of temporal phenomena during the "Prognoz 9"
 mission and "GRIF" experiment on board "Mir" station}

\subsection{The Prognoz 9 and "Mir" OS GRIF experiments.}

Both, Prognoz-9 and GRIF experiments use wide-field, hard X-ray spectrometers, which 
provide long-term observations of periodic process sources. Although the observational conditions 
were specific in the each experiment, the main X-ray instruments as well as the technique of the 
periodic sources revealing were quite similar.
The observations of galactic sources in hard X-rays (10-200 keV) were made in 1983-84 
during the complex experiment on a high-apogee ($\sim 720000$ km) satellite Prognoz-9 with a 
wide field of view (FOV) ($\sim 45^o$ FWHM) scintillator spectrometer ($\sim 40 cm^2$ effective 
area) \citep{Kud85}. The X-ray instrument was installed in such a way that the 
center of its field of view, averaged over the satellite's rotation period ($\sim 120 s$), coincided 
with the spin axis which pointed in the solar direction every 5-7 days. According to the experiment 
conditions sky areas adjacent to the ecliptic plane ($\pm 25^o$ - for instrument beam FWHM) were 
observed and slow ($1^o/day$ on average) scanning along the ecliptic was made. The count rates 
for X-ray photons were measured over the energy ranges of 10-50, 25-50, 50-100, 100-200 keV.
The region of the sky that was observed during the experiment is shown in equatorial 
coordinates in Fig 1. The points of the sky toward which the satellite's axis was pointed at different 
times (the dates in the figure refer to the origin of the corresponding intervals of constant 
orientation) are also marked in this figure. The main X-ray transient pulsars and the Galactic equator 
are shown. During the observations from November 1983 to February 
1984 of the sky region near the Galactic Centre (outlined by the closed solid line in Fig. 1), the 
count rates were considerably higher than the background count rates. Since each source in that region can 
be observed as long as 100 days, while virtually continuous measurements of count rates averaged 
over 10 s were made, the experiment provided favorable conditions for the study of periodic events 
over a wide range of periods.
   \begin{figure}
   \centering
   \resizebox{\hsize}{!}{\includegraphics{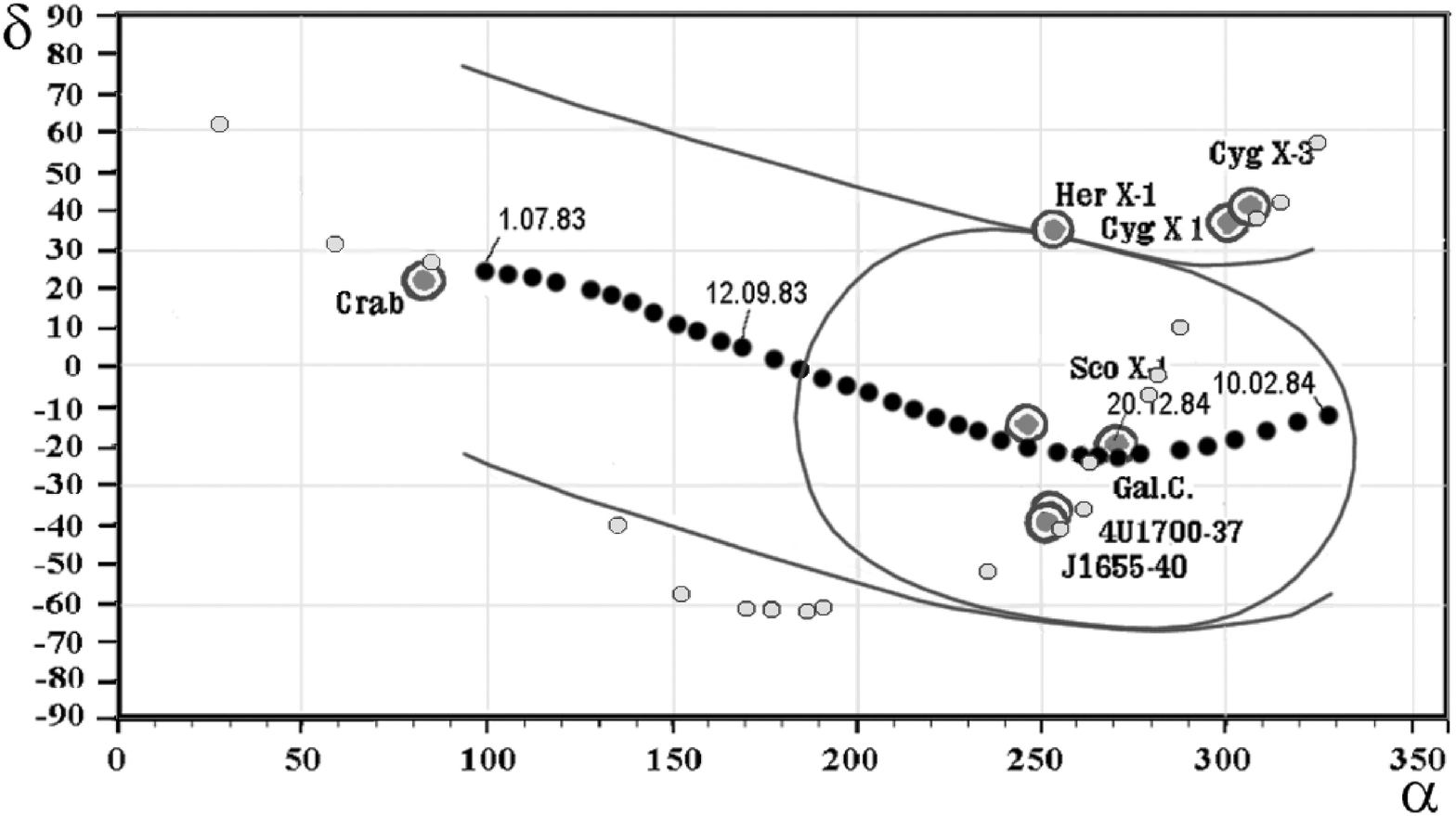}}
      \caption{ The region of the sky observed in the experiment 
                onboard Prognoz 9 satellite. 
              }
         \label{Fig1}
   \end{figure}

The long-term observations of the Galactic Center region as well as of some other sky 
regions were also conducted during the GRIF experiment on-board "Mir" orbital station (mean 
altitude $\sim 400$ km, orbit inclination $51^o$, orbital period $\sim 90$ min) from October, 1995 
to June, 1997 \citep{Kud98}. The scintillation spectrometer PX-2 for the energy range 
$\Delta E_{\gamma} = 10-300$ keV of detected photons with the effective area $S \sim 300 cm^2$, 
and field of view $\Omega \sim 1$ sr was the main instrument for astrophysical observations. It 
consists of 7 identical detector units of the "Prognoz-9" instrument type with crossed FOVs. The 
axes of these detector units were shifted on $5^o$ respectively to each other. This allowed us to 
observe almost the same area of the sky with all detectors simultaneously, and on the other hand, in 
the case of temporal phenomena registration to determine the source direction by the output data 
from each detector. The instrument provides flux measurements in energy ranges: 10-50, 25-50, 50-
100, 100-200 and 200-300 keV. Information was transmitted to the Earth in 16h-long sessions of 
continuous observations; the interval between them typically ranged from several hours to several 
days. A total $\sim 200$ sessions were conducted during the experiment, from which $\sim 150$ 
without many telemetry failures and incorrect times of output data recordings were chosen for 
subsequent analysis.
Due to the rigid fastening of the detector units to the station instrument panel its orientation 
was determined by the station orientation mainly in two modes: 3-axes stabilization and orbital. In 
the first case the instrument axis had fixed orientation in space while in the second case it slowly 
($\sim 4^o/min$) scanned the sky by the station orbital motion. Thus the different parts of the sky 
including Galactic Centre and Anti-Centre regions were accessible for observations during this 
experiment. 
   \begin{figure}
   \centering
   \resizebox{\hsize}{!}{\includegraphics{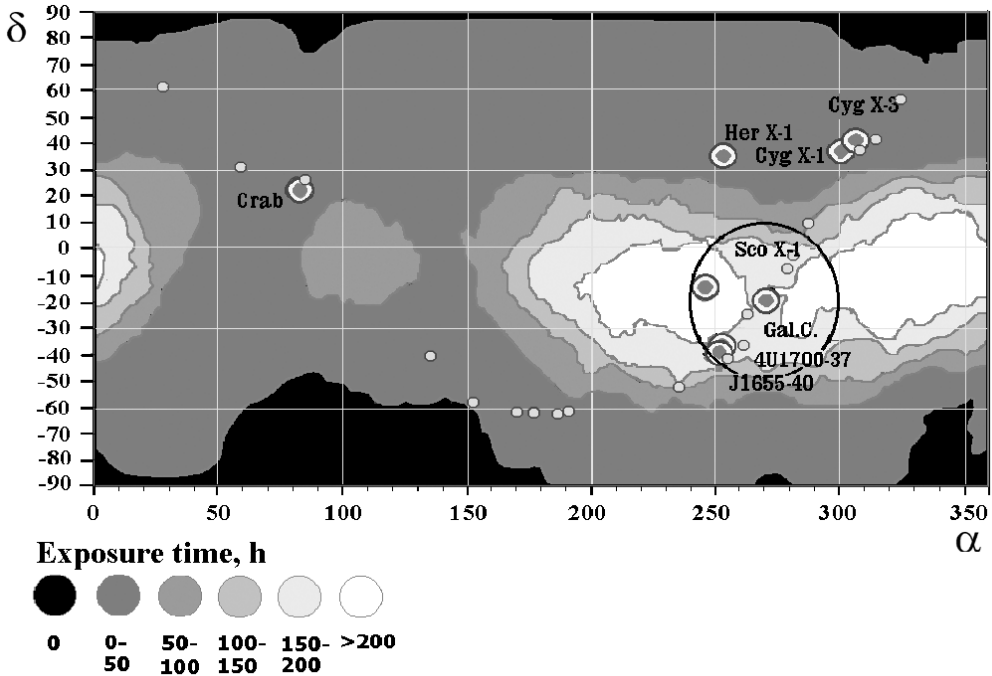}}   
      \caption{ The region of the sky observed in the GRIF experiment 
                onboard "Mir" orbital station. 
              }
         \label{Fig2}
   \end{figure}

	The sky region observed during the experiment is shown in Fig. 2 in equatorial coordinates. 
Different shades of gray represent the exposure time throughout the entire experiment. 
When estimating the exposure time, we 
assumed the angle between the source direction and the PX-2 axis to be no larger than $30^o$. In 
addition, we exclude the "Mir" residence time in the regions of trapped radiation. As a condition for 
the source being not shadowed by the Earth, we considered the requirement of PX-2 orientation to 
the sky, i.e. the angle between the PX-2 axis and the nadir-zenith direction should be within the 
range $0^o-90^o$. The figure also shows the brightest X-ray sources, slow pulsars and the Galactic 
equator. The typical exposure time can be determined by using observing conditions for the Galactic Center. 
Figure 2 shows a circumference with a radius of $30^o$ whose center coincides with the Galactic 
Center. There are sources within the region in the sky bounded by this circumference during the 
observations of which the PX-2 effective area accounted for no less than $50\%$ of its geometric 
area. As one can see from the figure, the total observing time of the Galactic Centre with $\geq 50\%$ 
efficiency was $\sim 200$ h. The exposure times of other Galactic sources (for example, 4U1700-37) 
are similar.

\subsection{The data processing technique}
	To reveal the periodic processes the time sets of X-ray instrument outputs were analyzed. In 
the case of the Prognoz 9 experiment such primary outputs were the mean count rates for 10 s. In 
the case of "Mir" GRIF experiment primary outputs were the mean count rates for 5 s. These time 
sets were subject to random and regular variations, which produce the 
background for periodicities. These include rises in the X-ray flux due to solar flares and cosmic 
gamma-ray bursts, changes of the background count rate in the X-ray channels caused by variations 
of charged particle fluxes that bombarded the spacecraft, and some others. The counts in X-ray 
channels can be caused partially by the stochastic variations in the total intensity of the emission 
from the sources, which were simultaneously within the instrument's field of view. Both, Prognoz 9 
and "Mir" GRIF experiments were capable of eliminating the effect of some background factors 
with the use of direct measurements of the individual background components.
	Significant sporadic rises in the count rates in the X-ray channels were mainly attributed to 
hard X-ray bursts from solar flares and cosmic gamma-ray bursts. Since the latter were recorded 
during the experiments rather rarely \citep{Kud88,Kud02}, 
they could not significantly affect the background in the search for periodicities. In the Prognoz 9 
experiment the Sun was constantly within instrument's field of view, thus its X-ray activity was 
monitored continuously, and about 800 solar X-ray bursts were detected \citep{Abr88}. 
To identify such bursts the data from the Prognoz 9 RF experiment \citep{Val79}, during 
which the Sun was continuously monitored in the band 2-10 keV by an instrument with a small 
($<10^o$) field of view, as well as the data of ground-based optical and radio observations of the 
Sun \citep{Cof84} were used. In the search for periodicities the time intervals during which the 
solar bursts were detected were excluded from analysis. Since such bursts are rather short-lived, the 
total duration of the time intervals rejected in this way in Prognoz 9 experiment was short, no longer 
than 0.3$\%$ of the total exposure time, for example, for the observations of the Galactic-Centre 
region. During the OS "Mir" GRIF experiment solar X-ray bursts were not detected at all because of 
the low solar activity during that time (1995-1997) \citep{Cof01}.

	To remove those variations in the Prognoz 9 experiment outputs which are correlated with the 
variations of the charged particle fluxes the regression analysis of the count rates in the X-ray channels 
($N_x$) and in the charged-particle channel ($N_z$) was used. The relation between the count rate 
in a given X-ray channel and the count rate of charged particles detected by the anticoincidence cap 
can be assumed to be linear. Thus, the initial count rates in the analyzed time series can be 
represented as a superposition of the count rates that characterize the photon flux $\sim N_{x}^*$ 
under study and the additional count rates due to the charged particles ($\alpha N_z$):
\begin{equation}
N_x = N_{x}^{*} + \alpha N_{z}.
\end{equation}
The linear regression coefficients $\alpha$ were determined from the sufficiently long ($\sim 100$ 
days) time series that corresponded to the regions of the sky under consideration. The most 
significant linear regression coefficients were obtained for the channels 50-100 and 100- 200 keV 
($\alpha \sim 0.04$ photon per particle). The effect of variations in the flux of charged particles in 
the channels 10-50 and 25-50 keV turned out to be weak. The $\alpha$ coefficient values were used 
to obtain the time series of count rates $N_{x}^{*} = N_{x} - \alpha N_{z}$ for subsequent 
analysis.

	One of the main peculiarities of the GRIF experiment was the possibility of simultaneously 
monitoring all the principal components of background-producing emissions in the near-Earth space 
on "Mir" station orbits. Thus, the large-volume CsI(Tl) scintillator detectors of NEGA-1 instrument 
independently detected the local gamma-quanta ($\Delta E_{\gamma} = 0.15-50$ MeV, 
$S_{\gamma} \sim 250 cm^2$) and neutrons ($\Delta E_{n} > 20$ MeV, $S_{n} \sim 20 cm^2$) 
produced by the interaction of cosmic rays with the spacecraft material and the Earth's atmosphere. 
The FON-1 electron detector (($\Delta E_{e} = 40-500$ keV) of high sensitivity was used to check 
the sporadic increases in X-ray flux attributable to bremsstrahlung from precipitating energetic 
magnetosphere electrons, which could simulate astrophysical phenomena (gamma-ray bursts, 
transients). Due to a large geometric factor ($\Gamma \sim 80 cm^{2}sr$) it could detect even 
relatively small electron fluxes outside the zones of captured radiation. The FON-2 charged-particle 
detector ($\Delta E_{e} = 0.04-1.5$ MeV, $\Delta E_{p} = 2-200$ MeV), which was free from 
overloading in the radiation belts because of its small geometric factor ($\Gamma \sim 0.5 cm^2sr$), 
was used for background measurements when the "Mir" station crossed the South-Atlantic anomaly 
and spurs of the outer radiation belt.

	Since the relatively large orbital inclination and periodic crossings of the zones of captured 
radiation the PX-2 instrument background count rates underwent variations. However, the 
combination of active and passive shields greatly reduced the background variations, in particular, 
the latitudinal variations in the main X-ray channels 25-50 and 50-100 keV. To extend the 
sensitivity range of the instrument the latitudinal count rate variations in high energy photon 
channels (100-200, 200-300 keV) were removed using regression analysis of the PX-2 X-ray 
channel outputs ($N_x$) and the NEGA-1 gamma-quanta channel outputs ($N_{\gamma}$). Since 
the NEGA-1 detectors were inside the OS "Mir" orbital module, they detected mainly the local 
gamma-emission. The additional count rate in a given X-ray channel attributable to the local 
emission may be assumed to depend linearly on the count rate of gamma-quanta detected by NEGA-
1. In this case, the initial count rates in the analyzed time series $N_x$ can be represented as a 
superposition of the X-ray count rate proper $N_{x}^*$, which characterizes the photon flux under 
study, and the additional count rate $\alpha N_{\gamma}$ attributable to the local gamma-
emission:
\begin{equation}
N_{x} = N_{x}^{*} + \alpha N_{\gamma}.
\end{equation}
The linear regression coefficients $\alpha$ were determined over the entire observation interval when there 
were no bright sources of hard emission within the PX-2 field of view. The $\alpha$ coefficient 
values were used to obtain the time series of count rates $N_{x}^{*} = N_{x} - \alpha 
N_{\gamma}$ for the subsequent analysis. The suppression of background variations with the use 
of readings in the 150-500 keV gamma-quanta channel yielded the most significant result. After 
applying the regression procedure, the residual variations in the X-ray channels attributable to the 
latitudinal variations accounted for no more than $\sim 3\%$ of the corresponding means, which is 
several times less than the expected amplitude of the variations attributable to the emission from the 
most intense Galactic source.

	In the search for periodicities, the time series of count rates (which were cleaned from solar 
bursts and background variations caused by charged particles) were processed by the standard 
epoch-folding technique \citep[see, e.g.][]{Ter92}, which was modified to accommodate the 
specific features of the Prognoz 9 and GRIF data. The intervals of observations were broken up into 
segments with duration equal to the trial period under consideration. The sequences of count rates 
that corresponded to these segments were added together, and an average phase dependence of the 
count rate was constructed for this trial period. The amplitude of the periodicity corresponding to 
the trial period (the actual or randomly simulated one) can be described by the rms deviation 
(${\sigma}^2$) of the numbers $M_i$ that constitute the average phase dependence:
\begin{equation}
\sigma^{2} ={\sum\limits_{i=1}^k (M_i - \bar{M}) \over {k-a}}.
\end{equation}
Here $\bar{M}$ is the mean count rate determined from the entire analyzed time series, $k = {T \over 
\Delta T}$, where $\Delta T$ is the bin duration of the count rates that constitute the mean phase 
profile; and $T$ is the trial period. If the periodicity related to the period under study contains both 
a real periodic component and a noise component, then because of their independence, ${\sigma}^2$ 
can be represented as the superposition:
\begin{equation}
\sigma^2 = \sigma^{2}_{pp} + \sigma^{2}_{noise},
\end{equation}
where $\sigma^{2}_{pp}$ and $\sigma^{2}_{noise}$ characterize the amplitudes of the 
corresponding components. 
In general, the dependence $\sigma^{2}(T)$ (periodogram) can be represented as the 
superposition of a noise continuum (ideally, a smooth function of $T$) and discrete peaks of the 
existing periodicities that correspond to the main period and its multiplies. The continuum in the 
periodogram is determined by a superposition of different (noise) components. One of such noise 
factors, shot noise ($\sigma^{2}_{shot}$), is related to the finite number of photons with a Poisson 
distribution recorded per bin. This component in the periodogram can be calculated from the mean 
count rates in the channels; it is linear for periodograms with a constant bin length, 
\begin{equation}
\sigma^{2}_{shot}(T) = \bar{N} \cdot {1\over\Delta} \cdot {T\over T_{max}},
\end{equation}
where $\bar{N}$ is the mean count rate, $\Delta$ is the bin length. Clearly, this component is always 
present in the noise, and the observed continuum can not be lower. The excess over the "shot" 
component in the total noise continuum can be explained by non-periodic variations in the fluxes 
from the observed X-ray sources (Galactic noise) and by variations in the instrumental background. 
For a wide range of periods, the results of data processing by the epoch-folding technique 
are more conveniently presented by plotting the inverse period (frequency) along the x-axis and by 
specifying a linear frequency grid in the calculations. In this case, the absolute FWHM of the peaks 
corresponding to periodicities with similar shapes is the same in the frequency spectrum at any 
periods. If we plot the parameter ${\sigma^2} \over T$ ($T$ is the period in bins) along the $y$
axis, the frequency continua of noise components correspond, to an accuracy of a factor, to those of 
a power spectrum in a Fourier analysis. In particular, the quantities that characterize shot noise are 
constant at all frequencies (white noise). The scatter of points in the real spectrum about the 
frequency-averaged values of ${\sigma^2} \over T$ in the noise segments in the vicinity of the peak 
under consideration gives the variance $\sigma_{\sigma}$, which can be used to estimate the 
significance of the periodicity corresponding to this peak (in the case of Prognoz 9 data 
$\sigma_{\sigma}$, values were calculated with the use of $\pm50\cdot\Delta T_{FWHM}$  relative to the peak,
where $\Delta T_{FWHM}$ is the mean width of the peak on the periodogram; for 
these intervals, the background level between peaks was essentially constant). Since the determining 
of the mean values of ${\sigma^2} \over T$ was performed by averaging over a large number of 
points in the spectrum, the error of the mean itself can be ignored, while the significance of the 
identification of a periodicity is determined by the ratio ${\sigma_p/\sigma_{\sigma}}$.

\section{Results of x-ray transient pulsars observations in the 
Prognoz 9 and GRIF experiments}

\subsection{The Transient Pulsars under Consideration and the Criterion of the Detection of Significant Flux.}

Using the method described above, all the data obtained in the Prognoz 9 and GRIF 
"Mir" experiments were analyzed. To select the significant periodicities the condition that the 
amplitude of the peak corresponding to the main period should exceed the mean noise value 
${\sigma^2}_{noise} \over T$ by more than $k\sigma_{\sigma}$ was established. For each 
observation the real amplitude distribution of peaks in the search interval was analyzed (amplitude 
of points was measured in $\sigma_{\sigma}$ units). Thus the level of significance was chosen 
according to the characteristics of peak amplitude distribution individually for each frequency 
spectrum. The reason of such approach is that the nature of a significant part of background on 
frequency spectra is not stochastic. It is connected with real processes in other frequency range or 
with peculiarities of used periodogram method described above. 
To search the periodic emission from X-ray pulsars in Prognoz 9 and GRIF "Mir" 
experiments it is necessary to analyze the frequency spectra in the range of small periods 
comparable with the duration of one telemetric data output record. In such a way the number of bins 
in trial period $T$ was chosen constant and equal 20 only for the large period values (more than 
$20\Delta t$, where $\Delta t$ is the time of one output record), while for periods shorter than 
$20\Delta t$ the number of bin was chosen equal to the ratio $T\over \Delta t$ rounded off to the 
nearest integer. The mean phase profiles constructed with such dividing on bins allow to obtain the 
pulsation profile in details and provide the sufficient elimination of random component. The 
minimal trial period for all periodograms and frequency spectra was equal $5\Delta t$.
The analyzing time sets are the numbers of count rate values averaging actually for the time 
from the beginning to the end of an output telemetric record. These count rate values correspond to  
all intermediate time moments of the record but not only to one of them. Thus, to obtain the mean 
phase profile in the case, when the bin boundary lays in the output record time interval, we add to 
the summing up counts in each bin the mean count in given output record with the weight 
proportional to the time of that part of record.

The unavoidable features of frequency spectra in the range of small periods several times 
greater than the time between the points of time series ($\Delta t$) is caused by the Naykvest noise 
number of narrow peaks with periods $\Delta t \cdot(n+1/2)$, $\Delta t \cdot(n+1/3)$, $\Delta t 
\cdot(n+2/3)$, $\Delta t \cdot(n+1/4)$, …where n is integer ($n = 5, 6, ...$). The nature of these 
peaks is connected with the averaging technique by the mean phase profile construction. Such peaks 
were excluded from the frequency spectra under their consideration. All frequency spectrum 
points corresponding to the periods in the limits $\pm0.01 \cdot\Delta t$ from the periods multiple 
to $\Delta t, \Delta t \cdot1/2, \Delta t \cdot 1/3, \Delta t \cdot 1/4$ were also excluded. The total 
length of the all excluded parts of frequency spectrum is negligible. 

Another background peak series appear on the frequency spectra because of the spacecraft 
rotation. In the case of Prognoz 9 data (Prognoz 9 satellite rotated with period about 2 min) a peak 
with period ~120 s and a number its multiples appear. Except for these series impair the analyzed 
spectra themselves the rotation of the instrument increase total variability resulting in increase of 
value of peaks multiple to the time of an output record. To clean the primary time set from 
variations caused by the satellite rotation the "whitening" procedure was used. Before the analysis 
those data time sets were identified, for which period and phase of satellite rotation were constant. 
Usually the time of such intervals was about 1 day. The frequency spectrum in the vicinity of the 
expected satellite rotation period peak was calculated for each of those intervals. From this spectrum 
the accurate satellite rotation period was determined. The mean phase profile corresponding to that 
period value was also obtained. Then after some smoothing such phase profile many times repeated 
with the phase conservation was subtracted from analyzed data series. After that the satellite 
rotation peak amplitude was at the level of the frequency spectrum points random dispersion.
As it was mentioned above count rates were measured in the Prognoz 9 experiment every 10 
s. In the GRIF experiment this time was 5 s. According to this, X-ray pulsars with pulsation period 
in the range 30-1500 s were considered. To the present time about 20 X-ray pulsars are known with 
periods in this range \citep{Bil97}. All of them are accretion-powered pulsars in binary 
systems including several Be transients. 

The time intervals during which these objects were in the FOV of the X-ray spectrometer 
PX-2 – the main astrophysical instrument of the GRIF experiment were determined at first. Because 
the full FOV of PX-2 instrument was $\pm 45^o$ from its axis, let us decide that given source is in 
the instrument's FOV if $\theta <30^o$, where $\theta$ is the angle between instrument axis and the 
direction toward the source. Some aspects of GRIF experiment described above leads to that time 
set of instrument output data obtained for given pulsar according to the condition $\theta <30^o$ was 
usually discontinuous. The parts of such time set when the source was in the FOV limits alternate 
with the time intervals, when the source was shadowed by the Earth, or the instrument was in the 
radiation belt or redirected. The frequency spectra restored from such "bit-continuous" data change 
strongly from point to point like a "saw". Without the separate peak corresponding to the periodic 
process there will be a "comb" of nearest peaks on the periodogram or frequency spectrum if they 
were obtained from data separated by large time intervals. It makes difficult to search pulsar 
periodicities if the pulsation period is not known with enough accuracy. Thus, further analysis was 
made only for the pulsars, for which the value of pulsation period in different epoch of observations 
could be estimated basing on the world data about the drift of their period.
Taking into account the BATSE CGRO and some other experiment data (Bildsten et al., 
1997) 7 transient pulsars were chosen, for which both criteria as on the pulsation period, as on the 
$\theta$ angle were valid. For these pulsars the time intervals $[T_{min}, T_{max}]$ in which their 
pulsation periods should lay in the GRIF observations epoch were chosen.

The same criterion was 
used for Prognoz 9 experiment: the $\theta$ angle between the instrument axis and direction toward 
the source was defined in different time intervals for each of known transient pulsars with 
appropriate pulsation period, then only objects for which $\theta <30^o$ were considered. Only 3 
transient pulsars appeared in the "Prognoz 9" field of view not more than about $30^o$ far from the 
instrument's axis. They are also listed in the GRIF catalogue. The interval $[T_{min}, T_{max}]$ of 
periods for the search for pulsations from A0535+26 was chosen by the extrapolation of its period 
value to Prognoz 9 epoch. In the case of Sct X-1 and GS1722-36 pulsars in view of insufficient data 
about drift of their periods the interval of search for periodicities was chosen about 1\% of 
corresponding periods measured in epoch when they were discovered. 
The best sensitivity in both experiments was achieved in the channels corresponded to 
energy ranges 10-50, 25-50 keV (the typical energy spectra of transient pulsars were taking into 
account also). Thus, just these ranges were chosen for primary search of pulsation. The data in other 
ranges were analyzed in case of revealing of significant peaks on the frequency spectra in the 
channels 10-50 and 25-50 keV. 

By the analysis of GRIF experiment data the time intervals containing the sufficient enough 
number of output readings were chosen for each of considered pulsars. Typically they were several 
consequent seances each lasting for about 16 hours with gaps between them. Besides such rather 
comfortable intervals the analysis was made also for more long ones. The time of analyzed intervals 
was increased for those pulsars, which were in the instrument FOV rarely and during a short time. 
The time sets obtained as the result of primary data processing in the ranges 10-50, 25-50, 50-100, 
100-200 keV were analyzed by the epoch folding technique, which was described above in details. 
As the result the periodograms (frequency spectra) have been obtained for the range of periods 
30 – 1.5 $\cdot 10^3 s$ with the trial period net uniform on the inverse period value, which includes 
500000 points. Such detail is sufficient enough to resolve the peaks corresponding to the real 
pulsation. The view of the all set helps to interpret correctly the frequency spectrum peculiarities 
connected with the satellite orbit period harmonics. This is especially important in the case of study 
of pulsars with long (more than 200 s) periods. 

To reveal the periodic pulsation on the obtained periodograms the software providing the 
subtraction of stochastic component and allowing evaluate the peak significance was elaborated. 
With the use of this software the periodograms were transformed into a form trial period, $T$ (or its 
inverse value – frequency, $T^{-1}$) versus the number of standard derivation characterized the 
dispersion of points on the primary periodogram near the analyzed peak, $\sigma_{\sigma}$. To 
elaborate the criterion of significance of separate peak on the periodograms the dispersion 
$\sigma_{\sigma}$ was analyzed. The $\sigma_{\sigma} - T^{-1}$ representation allows to 
determine the significance criterion uniformly on the all spectrum. To exclude the influence of non-
stochastic peaks (of astrophysical origin or imitated due to the some methodic reasons) on the mean 
and the standard deviation values they were calculated from the $\pm5000$ points on the frequency 
spectrum in two steps. 
On the first step all points where the peak amplitudes were higher than four standard 
deviations ($4\sigma$) were excluded. 

For the estimation of the peak amplitude significance level it 
was taken into account that the amplitude distribution of the points on the frequency spectrum is not 
normal. Moreover, it is determined by the several background factors with various contributions on 
the different intervals of observations as well as on the different parts of frequency spectrum. From 
this reason, the amplitude distributions of the points on the frequency spectra were evaluated 
empirically for each case. The parts of the frequency spectrum near the expected inverse pulsation 
period in the limits $\pm6\%$ from the corresponding value were used. As the result the empirical 
distributions characterizing the dispersion of points on the periodograms were obtained. With the 
help of these distributions the significance levels were estimated for each time intervals were the 
pulsation with corresponding periods could be observed. The exact values of threshold amplitude 
were chosen taking into account the form of each distribution. 

To reveal the periodicity of astrophysical origin the sufficiently hard condition was used: the 
probability of that at least one random peak on the interval of search exceed the chosen threshold 
should be less than $1\%$ with the real peak amplitude distribution. Obviously the threshold value 
depends not only on the stochastic process parameters but also on the width of the frequency range 
of the search, which could be expressed via the numbers of FWHM. If some peak on the range of 
search exceeded the threshold level, we can conclude that the corresponding X-ray intensity 
variation is not casual. In those cases when the maximal peak amplitude did not exceed the 
threshold, we decided that periodic process was not detected. The amplitude of maximal peak on the 
interval of search was used for estimation of the upper limit of pulsation component intensity. It 
means that in some cases it was not the absence of maximums on the frequency spectrum, but the 
absence of significant peaks on the frequency spectrum. 

The frequency range of search of pulsation was determined separately for each pulsar from 
the evaluation of the possible uncertainty of pulsation period due to its drift. In the most cases this 
range several hundreds times exceeded the experimental resolution of the period value. To analyze 
the dynamics of pulsation period the specific programs, allowing compare the periodic process 
intensity in the different interval of observations, were elaborated. As the result, the two-
dimensional diagrams reflected the dependencies on time as of period value as of periodic process 
intensity were obtained. In such a way the time variations of pulsation periods of those sources, 
which could be observable in the GRIF experiment, were examined. Additionally the search and 
timing of pulsation of the same sources with the use of Prognoz 9 data were made to study the 
pulsation period dynamics on the long-term database.

\subsection{The Prognoz 9 and GRIF "Mir" summary catalogue of observed transient pulsars}

The results of the search of periodic emission from X-ray transient pulsars in the GRIF Mir 
and Prognoz 9 data are presented in the Tables 1, 2. The periods corresponding to the maximal 
peaks on the frequency range of search as well as the pulsed fluxes (significant values or upper 
limits) obtained from those peak amplitudes are presented in the Table 1 for different intervals of 
observations in the GRIF experiment. It is necessary to note that searches of pulsation in the GRIF 
data were made only in the 25-50 keV energy range because it was the most pure from the 
background variations. For those transient pulsars which were observed in the GRIF experiment the 
mean total and pulsed fluxes obtained in the BATSE CGRO \citep{Har04, Bil97} and the 
Integral \citep{Lut04a,Rev04,Mol04} observations in the energy range nearest to the GRIF 
one were also added in the Table 1 to compare with GRIF data. The mean orbital phase values in the 
time of GRIF observation form the last column of the Table 1.

\begin{table*}
\caption{The results of the search for periodic emission from transient 
pulsars in the GRIF Mir experiment.}
\label{Table:1}
\begin{tabular}{l c c c c c c c c c}
\hline\hline
Pulsar & $T_{beg}-T_{end}$ & $T_{puls}$ & 
Flux $\cdot 10^{10}$ & Flux $\cdot 10^{10}$ & Flux $\cdot 10^{10}$ & 
Flux $\cdot 10^{10}$ & $T_{orb}$ & orbital \\
& & & Pulsed & Mean total & Pulsed & Total & & phase\\

& & & GRIF & BATSE & BATSE & INTEGRAL & & \\
& & & 25-50 keV & 20-40 keV & 20-40 keV & 20-60 keV  & & \\
\hline
	
& TJD & s & erg/$cm^2s$ & erg/$cm^2s$ & erg/$cm^2s$ 
& erg/$cm^2 s$ & d & \\
\hline

A0535+26 & 10091.5-10092.0 & 102.916 & 1.72 & 1.14$\pm$0.08 & 0-150 & & 110 & 0.144 \\
& 10218.2-10218.9 & 103.395 & 1.28 & & & & & 0.282 \\
& 10370.0-10460.0 & 102.935 & 0.689 & & & & & wide \\
& 10580.3-10580.9 & 103.491 & 2.84 & & & & & 0.533 \\
& 10614.4-10614.6 & 103.356 & 2.65 & & & & & 0.837 \\
4U1145-619 & 10092.8-10095.7 & 291.319 & 5.59 & 1.68$\pm$0.08 & 2-15 & & 187 & 0.56 \\
& 10385.0-10457.0 & 291.169 & 3.40 & & & & & 0.12-0.50 \\
& 10092.0-10581.0 & 291.252 & 4.46 & & & & & wide \\
A1118-615 & 10385.0-10454.0 & 405.540 & 8.03 & 0.174$\pm$0.06 & 4-30 & & & \\
& 10092.0-10581.0 & 406.123 & 6.54 & & & & & \\
EXO2030+37 & 10102.4-10105.4 & 41.643 & 2.44 & 0.129$\pm$0.053 & 2-12 & & 46.0 & 0.30-0.40 \\
& 10396.4-10477.0 & 41.644 & 1.43 & & & & & wide \\
Sct X-1 & 10072.7-10103.6 & 110.776 & 0.802 & n/d & & & & \\
& 10430.6-10442.4 & 111.391 & 1.04 & & & & & \\
& 10604.7-10615.0 & 110.452 & 1.26 & & & & & \\
EXO 1722-36 & 10050.0-10110.0 & 416.150 & 1.02 & & & 1.0052 & 9.7 & \\
& 10070.0-10090.0 & 415.696 & 1.09 & & & & & \\
& 10604.0-10616.0 & 415.810 & 1.56 & & & & & \\
& 10410.0-10415.0 & 414.195 & 0.941 & & & & & \\
SAX & 10028.6-10103.3 & 359.221 & 3.95 & 0.091$\pm$0.068 & & & & \\
 J2103.5+4545 & 10383.7-10474.7 & 360.818 & 2.82 & & & & & \\
& 10383.7-10580.9 & 360.818 & 2.74 & & & & & \\
& 10580.0-10581.0 & 357.137 & 1.18 & & & & & \\
IGR 16320-4751 & 10051.0-10052.6 & 1232.494 & 5.15 & & & 1.82 & & \\
& 10410.8-10414.4 & 1237.460 & 6.12 & & & & & \\
& 10434.6-10436.3 & 1170.022 & 3.00 & & & & & \\
& 10580.2-10580.8 & 1326.165 & 7.94 & & & & & \\
& 10609.7-10615.1 & 1249.065 & 2.30 & & & & & \\
IGR 16465-4507 & 10051.0-10052.6 & 223.900 & 3.17 & & & 1.232 & & \\
& 10410.8-10414.4 & 230.858 & 1.75 & & & & & \\
& 10434.6-10436.3 & 205.683 & 1.85 & & & & & \\
& 10580.2-10580.8 & 243.831 & 4.08 & & & & & \\
\hline
\end{tabular}
\end{table*}

	The same parameters as in the Table 1, but for the transient pulsars observed in the Prognoz 
9 experiment are presented in the Table 2.

\begin{table*}
\caption{The results of the search for periodic emission from transient 
pulsars in the Prognoz 9 experiment.}
\label{Table:2}
\begin{tabular}{l c c c c c c c c c c}
\hline\hline
Pulsar & $T_{beg} - T_{end}$ & $T_{puls}$ & 
Pulsed flux & Pulsed flux & $T_{orb}$ & orbital phase \\
& & & Prognoz 9 & Prognoz 9 & & \\
& & & 25-50 keV & 10-50 keV & & \\
\hline
& TJD & s & erg/$cm^2 s$ & erg/$cm^2 s$ & d & \\
\hline
A0535+26 & 5516.97-5522.73 & \bf{103.303} & \bf{1.61E-10} & \bf{3.61E-10} & 110 & 0.066-0.122 \\
& 5516.79-5518.70 & \bf{103.294} & \bf{1.76E-10} & \bf{6.48E-10} & & \\
& 5518.73-5522.73 & \bf{103.304} & \bf{1.05E-10} & \bf{3.07E-10} & & \\
& 5522.73-5528.74 & 103.124 & 1.14E-10 & 1.82E-10 & & \\
& 5528.74-5534.66 & 103.090 & 1.36E-10 & 3.21E-10 & & \\
Sct X-1 & 5681.49-5688.36 & 111.789 & 1.95E-10 & 3.01E-10 & & \\
& 5688.36-5695.40 & 110.673 & 1.20E-10 & 2.33E-10 & & \\
& 5695.40-5705.27 & 110.694 & 1.03E-10 & 1.70E-10 & & \\
& 5705.27-5711.27 & 111.743 & 1.29E-10 & 2.05E-10 & & \\
& 5681.49-5711.27 & 111.484 & 6.93E-11 & 1.40E-10 & & \\
EXO 1722-36 & 5668.40-5675.31 & \bf{411.147} & \bf{3.80E-10} & 6.38E-10 & 9.7 & \\
& 5675.31-5681.49 & 415.866 & 1.75E-10 & 3.18E-10 & & \\
& 5681.49-5688.36 & 417.425 & 1.18E-10 & 2.06E-10 & & \\
& 5688.36-5695.40 & 412.939 & 1.08E-10 & 1.84E-10 & & \\
& 5695.40-5705.27 & 415.379 & 1.03E-10 & 1.54E-10 & & \\
& 5668.40-5705.27 & 419.215 & 5.88E-11 & 1.03E-10 & & \\
\hline
\end{tabular}
\end{table*}

The values of the period of pulsations and the corresponding flux values are printed with 
bold font for the cases of significanf detection. As it can be seen from the Tables only two sources: 
A0535+26 and GS1722-36 demonstrated significant fluxes. The upper limits of pulsation intensity 
were obtained for other transient pulsars. The mean phase profiles concerning to those pulsation 
processes, which intensity exceeded the significance level, analyzed in details. The forms of mean 
pulsation profiles in different energy ranges were compared with results of other experiments - 
BATSE CGRO \citep{Bil97}, Ginga \citep{Nag89}. The properties of pulsation emission from 
A0535+26 and GS1722-36 observed in the Prognoz 9 experiment will be discussed in details below.

\subsection{A0535+26}

With the use of the presented above technique periodograms were obtained for the output 
count time sets in the ranges 10-50, 25-50, 50-100 keV, which were chosen for the intervals of 
observation when the Prognoz 9 X-ray instrument's axis was near the direction toward the location 
of the well-known X-ray transient pulsar A0535+26. Those periodograms are presented in Fig. 3a. 
The significant discrete peaks corresponding to the 103.3 s period are quite reveal on the 
periodograms. The value 103.3 s is equal to known A0535+26 pulsation period in the error 
limits caused mainly by the uncertainties due to the drift of period.
   \begin{figure*}
   \centering
   \includegraphics[width=12cm]{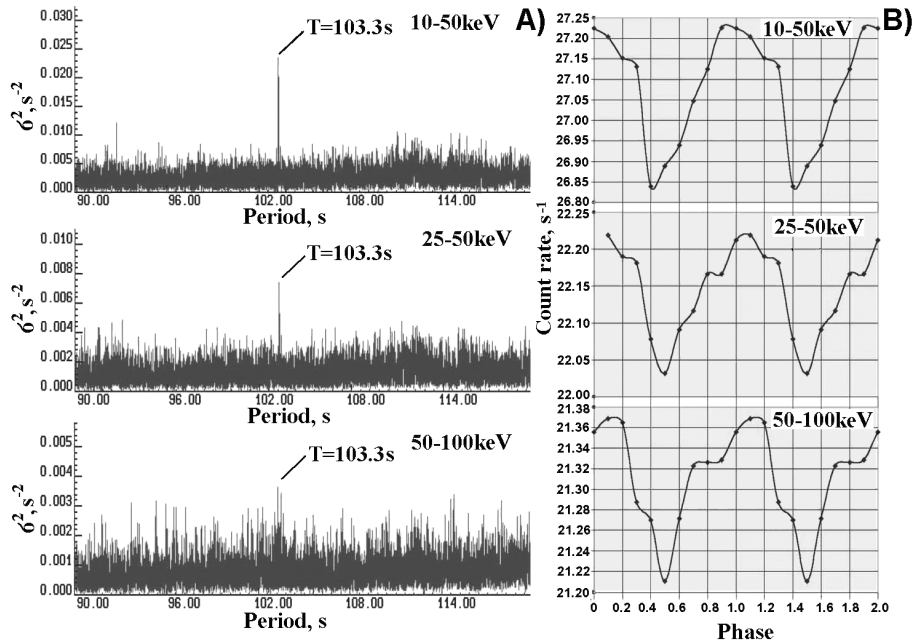}
      \caption{ A.The periodograms obtained by processing July, 1983 data
      of Prognoz 9 experimrnt in 10-50 keV, 25-50 keV and 50-100 keV energy
      channels. B.The mean phase profiles of 103.3 s pulsations from
      A0535+26 transient pulsar in correspondent energy ranges. 
              }
         \label{Fig3}
   \end{figure*}

The mean phase profiles of pulsation (light curves) were obtained for the most probable 
period equal to the value which was weight-averaged over the peak on the periodogram. The mean 
phase profiles of that period pulsation in the energy ranges 10-50, 25-50, 50-100 keV are presented 
in Fig 3b. In the range 100-200 keV pulsation was not revealed at the level of background 
variations. As it could be seen from the figures, the forms of presented light curves in the different 
energy ranges are similar and characterize the rather complicated mean pulsation profile. To the all 
profiles the narrow dip is typical. Its width decrease with increasing of the detected photon energy: 
about $20\%$ of the full phase in the 10-50 keV range, and about $10\%$ of the full phase in the 
50-100 keV range. Also some pequliarities probably being present on the phase profile near the 
$\sim 0.8, \sim 1.1, \sim1.3$ of the full phase are seen better in higher energy range.
In general the presented curves do not contradict the A0535+26 pulsation profiles measured in other 
experiments \citep{Nag89,Gio92,Bil97}.

The time intervals when the significant pulsation fluxes from A0535+26 were observed in 
the Prognoz 9 experiment presented in the Table 2 correspond to the orbital phases 0.066-0.122, i.e. 
the outburst decrease phase. Indeed, it was the July, 1983 pulsar outburst decreasing \citep{Sem90}.
The mean pulsation intensity values in the 25-50 keV energy range on the consequent time 
intervals during the Prognoz 9 observations are presented in the panel "a" of Fig 4. Those intensity 
values were obtained as the counts in the most probable peak at the period search interval on the 
periodograms folded over the corresponding intervals of observations. As it follows from the figure, 
despite the rather high error values the steady decreasing of pulsation intensity can be seen quite 
evidently. 
   \begin{figure}
   \centering
   \resizebox{\hsize}{!}{\includegraphics{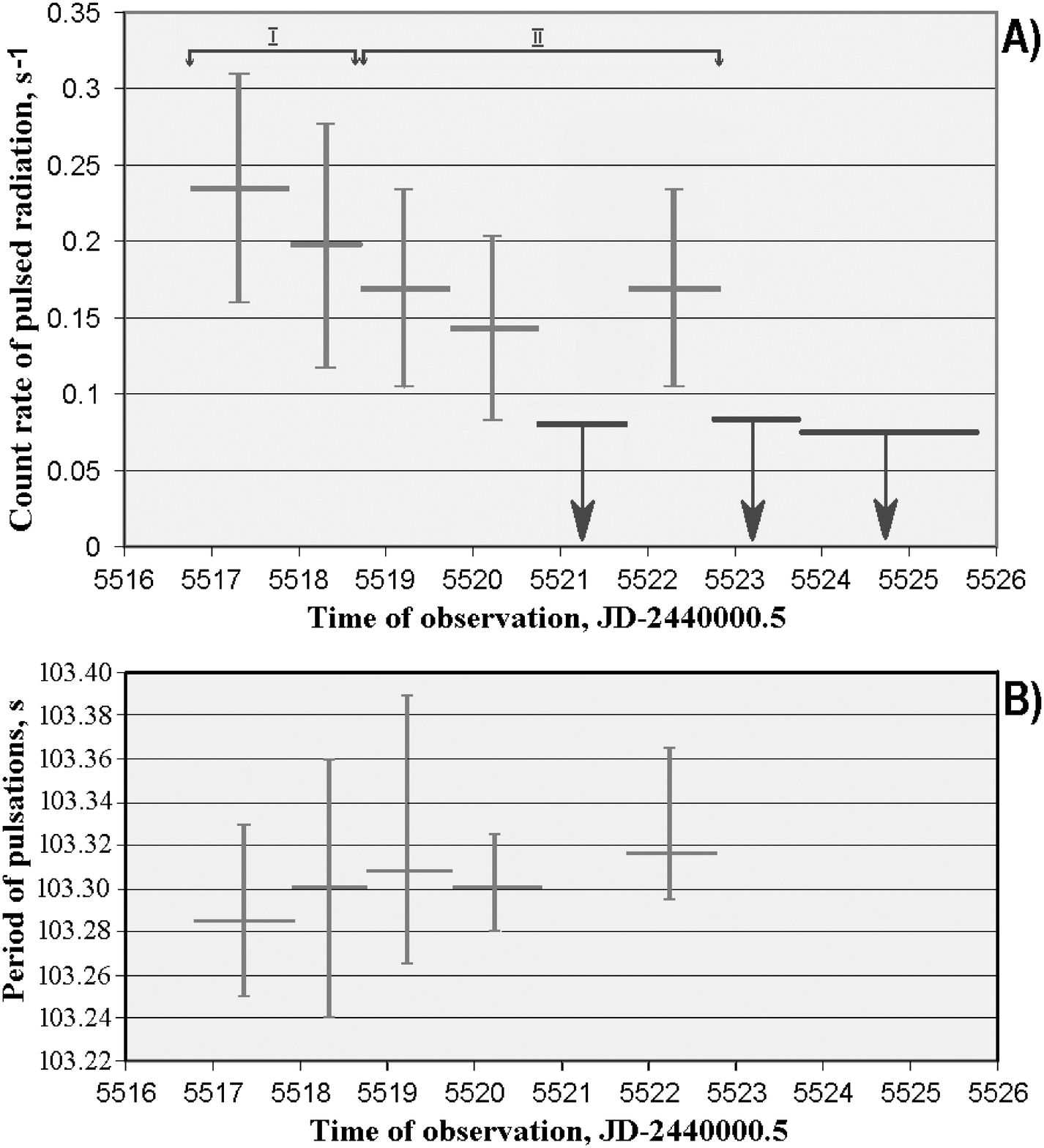}}
      \caption{ A.The mean pulsation intensity values in the 25-50 keV energy range on the consequent time 
intervals during the Prognoz 9 observations. B.The time dependence of the mean pulsation period obtained for 
those time intervals that pulsation intensity values.
              }
         \label{Fig4}
   \end{figure}

The time dependence of the mean pulsation period obtained for those time intervals that 
pulsation intensity values are presented on the panel "b" of Fig 4. The same values of the period of 
pulsations are presented on Fig 5 together with SMM data \citep{Sem90} corresponding to 
the primary stage of A0535+26 June 1983 outburst. We may conclude from this figure, that taking 
into account the limits of errors, that during the Prognoz 9 period of observation the period value 
has not changed significantly, so the acceleration of the NS rotation stopped at the final stage of the 
outburst. Moreover, quite evident trend of the period increasing as the intensity fall down can be 
noticed. This small period increasing does not contradict the correlation between spin-up rate and 
X-ray flux (and pulsed X-ray flux as well), which is well-known for many Be transient pulsars 
including A0535+262 \citep{Bil97, Fin96b}.
   \begin{figure}
   \centering
   \resizebox{\hsize}{!}{\includegraphics{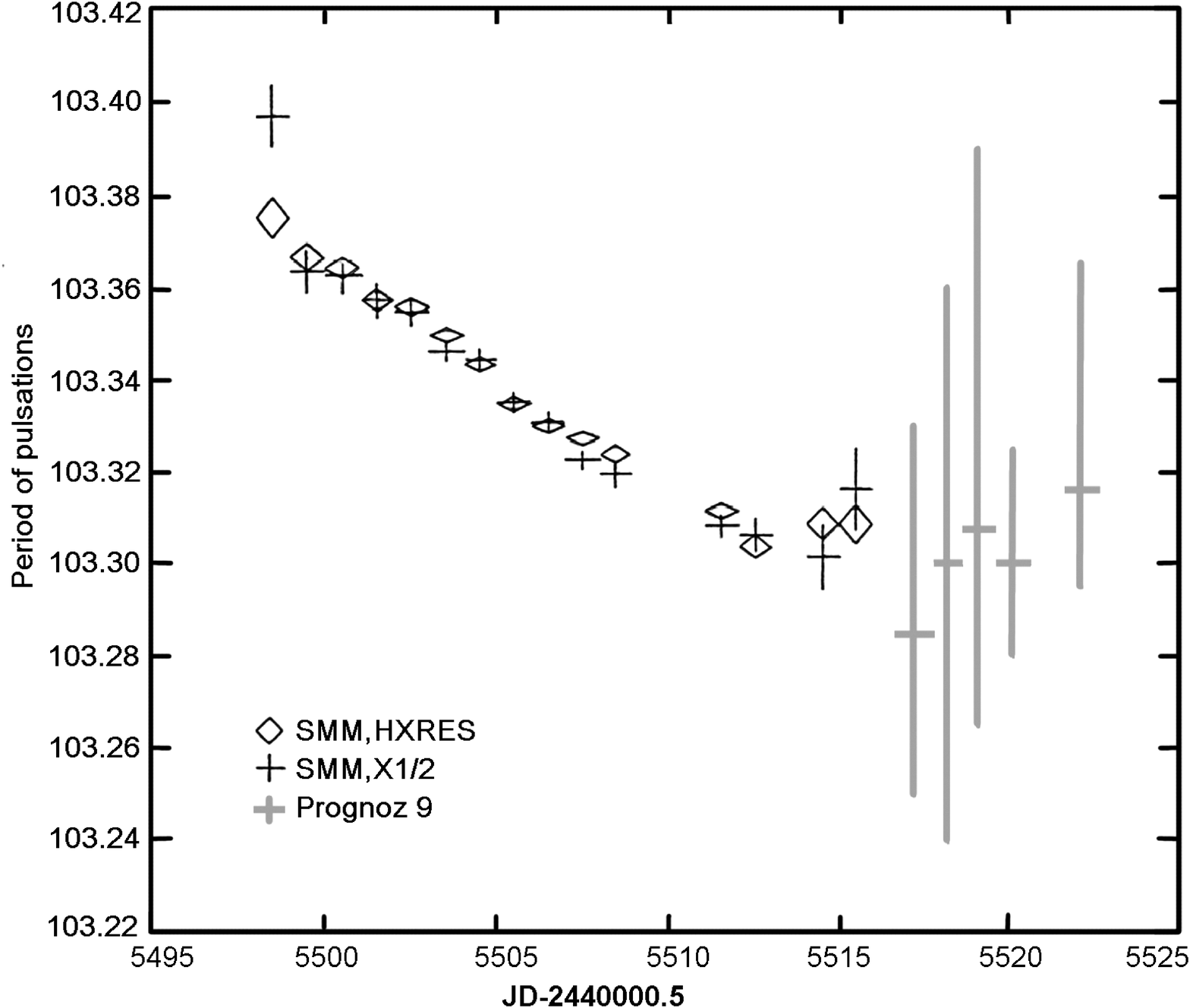}}
      \caption{ The time dependence of the mean pulsation period of A0535+26
      obtained in Prognoz 9 and SMM missions during June-July, 1983 outburst}
         \label{Fig5}
   \end{figure}

   \begin{figure}
   \centering
   \resizebox{\hsize}{!}{\includegraphics{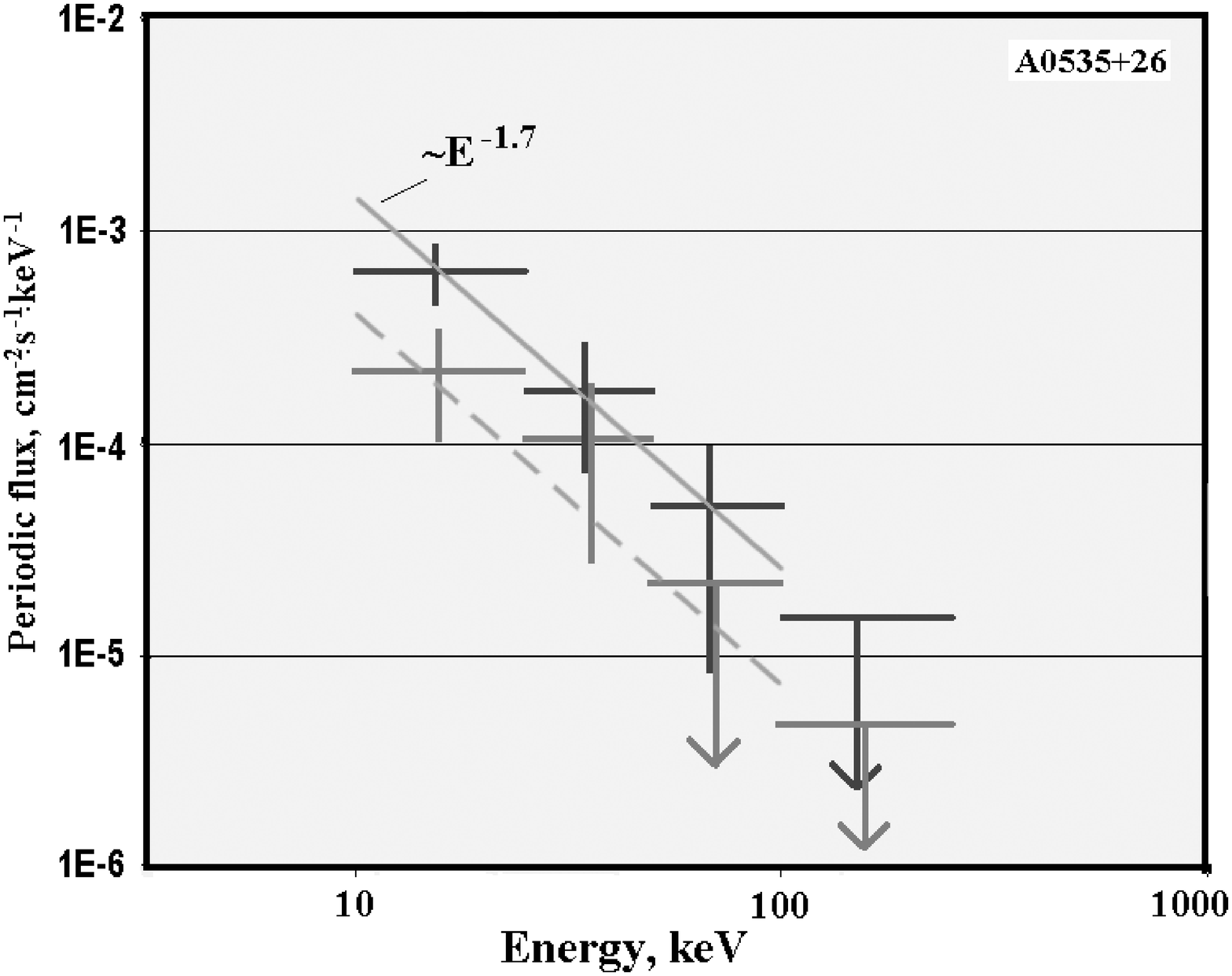}}
      \caption{ Energy spectra of A0535+26 pulsed radiation obtained in 
       Prognoz 9 experiment for TJD5516.8-5518.7 (upper) and TJD5518.7-5522.7
       (down) intervals of observation. Power low approximation with 
       slope $\sim 1.7$ is showen.}
         \label{Fig6}
   \end{figure}

The energy spectra of A0535+26 hard emission pulsation component, which were 
constructed for consequent time intervals marked on Fig. 4a as "I" and "II" are presented in Fig. 6. The best-
fit power approximation of the spectrum obtained for the interval "I", i.e. for the first half of 
observed intensity decreasing gives the slope $\sim 1.7$. As it could be seen from the Figure, such 
slope in the error limits also well approximates the spectrum for the second interval "II", when the 
pulsation intensity was sufficiently less. Thus, we may conclude that the energy spectrum of X-ray 
pulsation component does not changed essentially on the outburst-decreasing phase. 
The obtained energy spectra do not contradict the  "canonical" model for X-ray pulsars 
which is used often for the approximation of the full flux energy spectrum of transient pulsars \citep{Whi83} 
This three parameter approximation gives $I(E)=I_0\cdot E^{-\gamma}$ for low energy part $E<E_c$ and 
$I(E)=I_0\cdot E^{-\gamma}\cdot exp(-(E-E_c)/E_f)$ for $E>E_c$ 
The spectra obtained in Prognoz 9 experiment are 
satisfactory described by the "canonical" model using the parameters similar with ones of the full flux 
in 1984, april normal outburst of A0535+26
 \citep{Bor98}. There are photon index $\gamma\sim1.2$, cutoff energy $E_c\sim23$ keV and e-folding energy $E_f\sim19$ keV.
It indicates on that spectral hardness of the full flux and its pulsation component may be similar. 
Besides, there is no indication on the presence of any soft excess near the 10 keV. Such soft 
excess like a thermal or black body spectral components was observed in some binary X-ray pulsars 
including A0535+26 \citep{Muk05}. It was interpreted as the emission of a hot region 
near NS polar caps where accreting plasma deposits its energy. The absence of thermal component 
probably indicates on the very small radii of emitting region.

	Taking into account the probable distance to the A0535+26 source about 2 kps \citep{Stee98, Gia80}, 
we obtain the luminosity in the pulsed component 
$\sim1.5\cdot10^{35}$ erg/s, which corresponds to the measured pulsation flux 
$3.6\cdot 10^{-10} erg/cm^2\cdot s$ in the 10-50 keV energy range (see Table 2). The fraction of pulsed component is 
varied from $30\%$ to $70\%$ for typical A0535+26 outbursts \citep{Bor98}. Supposing that during Prognoz 9 mission 
just typical outburst was observed, we may estimate the full luminosity $\sim3\cdot10^{35}$. The comparison with 
the luminosity in the quiescent state $\sim3\cdot10^{33}$ erg/s (the BeppoSAX results in 2-20 keV range \citep{Muk05} ),
$\sim2.5\cdot10^{35}$ erg/s (EXOSAT measurements between outbursts in 1-20 keV range \citep{Mot91} ) and 
on the other hand with the luminocity of A0535+26 in "normal" outbursts $\sim10^{37}$ erg/s \citep{Bor98} 
allows to conclude that we 
have observed the final phase of the outburst. In favor of this guess the mentioned above absence of 
soft spectral component also may be caused by the squeezing of emitting region due 
to the accretion exhaustion. Nevertheless, the quite evident revealing of pulsation on the 
outburst-decreasing phase indicates on that considerable fraction of accretion disk material continued to fall 
onto NS surface, or centrifugally inhibited regime was achieved. 

	As the result of A0535+26 observations in the "Mir" GRIF experiment the only upper limits 
on the pulsation flux in 25-50 keV were obtained (see table 1). This source was observed during 
five time intervals. In two of them (10580.3–10580.9, 10614.4-10614.6 TJD) the orbital phase was 
more than 0.5, i.e. it was not any condition for outburst rise. Two intervals (10091.5-10092.0, 
10218.2-10218.9 TJD) corresponds the orbital phase 0.144, 0.282, i.e. principally the outburst finale 
stage may be found. However, the obtained upper limits ($(1.3-1.7)\cdot 10^{-10} erg/cm^2\cdot 
s$) are of the same order than the pulsation fluxes at the same energy range, which was observed in 
the Prognoz 9 experiment for the "joined" orbital phases ($<0.122$). Thus, it is quite possible that on 
the orbital phases $>0.14$ the pulsation fluxes really less than those limits, which are presented in 
Table 1. As for the last interval (10370-10460 TJD) it was too wide to reveal significant pulsation 
from any outburst phase on that rather intensive X-ray background variations in the GRIF 
experiment.

\subsection{GS1722-36}

The significant peak at the 411.7 s period, which is known as the pulsation period of the 
transient pulsar GS1722-36 (EXO1722-36), was revealed on the periodogram in the energy range 
25-50 keV. The periodogram was obtained for the time interval (5668.40 – 5675.31 TJD) of the 
most favorable observational conditions of this source in the Prognoz 9 experiment: the angle 
between the X-ray instrument's axis and the direction toward the pulsar was no more than $20^o$. 
The mean phase profiles corresponding to the 411.7 s period in different energy ranges are 
presented in Fig. 7a. For comparison the pulsation profiles obtained as the result of the Ginga 
observations of the GS1722-36 \citep{Nag89} are also presented in Fig. 7b. As it could be seen from the figures, 
in 10-50 and 50-100 keV energy ranges the 411.7 s periodicity is characterized by the single-
maximum mean phase profile. The form of these profiles do not contradict to the Ginga data, 
according to which the 19-38 keV profile is quasi-harmonic while at the lower energies it could be 
characterized by the evident twin peaks form with the second maximum sufficiently more intensive 
than first one. Although the Prognoz 9 profile significance in the 50-100 keV energy is not high 
enough to make some conclusions about its form in details, it is quite evident that maximum  
becomes narrower at higher energies: it is about a half of period in the 10-50 keV energy range and 
about $25-30\%$ of the period in the 50-100 keV. The 411.7 s periodicity in the energy range 100-
200 keV was not revealed with sufficient significance, thus the corresponding profile was not 
presented in the Figure. 

   \begin{figure}
   \centering
   \resizebox{6cm}{!}{\includegraphics{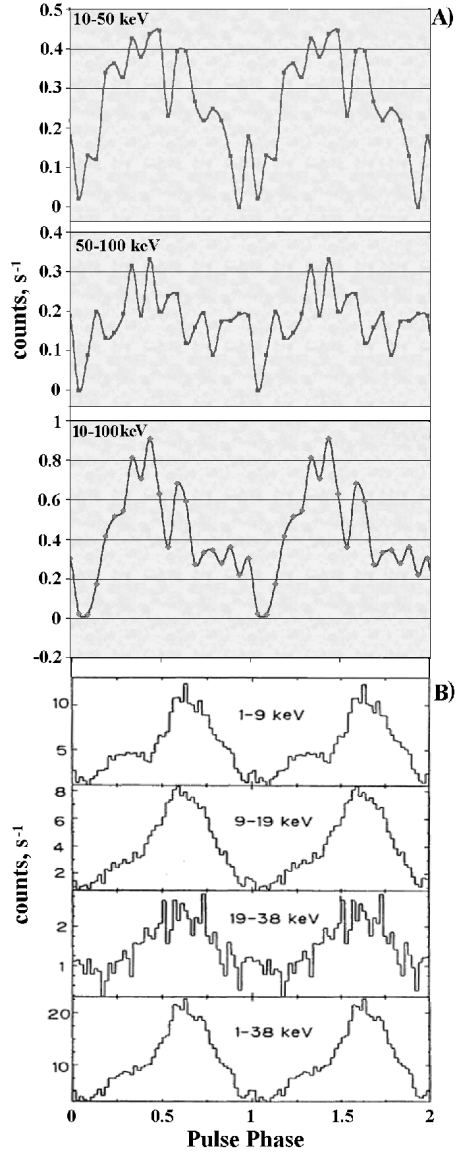}}
      \caption{ A.Mean phase dependencies obtained for GS1722-36 pulsations in
      several energy ranges. B. The Phase dependencies of GS1722-36 X-ray pulsations 
	obtained in Ginga experiment \citep{Nag89}. }
         \label{Fig7}
   \end{figure}

The energy spectrum of GS1722-36 pulsations is presented in Fig. 8. This spectrum 
corresponds to the pulsation intensities obtained from the peaks revealed on the periodograms 
calculated for the same time interval (5668.40 - 5675.31 TJD)in different energy channels. The best-fit power 
law approximation of the spectrum gives the slope $\sim 1.2$. Thus this spectrum is sufficiently harder than 
the typical spectrum of the most of transient X-ray pulsars \citep{Bil97}. Even taking into account that 
higher energy range used for power law approximation in BATSE experiment leads to the increase of estimated 
power index one may conclude that the obtained spectrum of pulsations is harder than typical total one.

   \begin{figure}
   \centering
   \resizebox{\hsize}{!}{\includegraphics{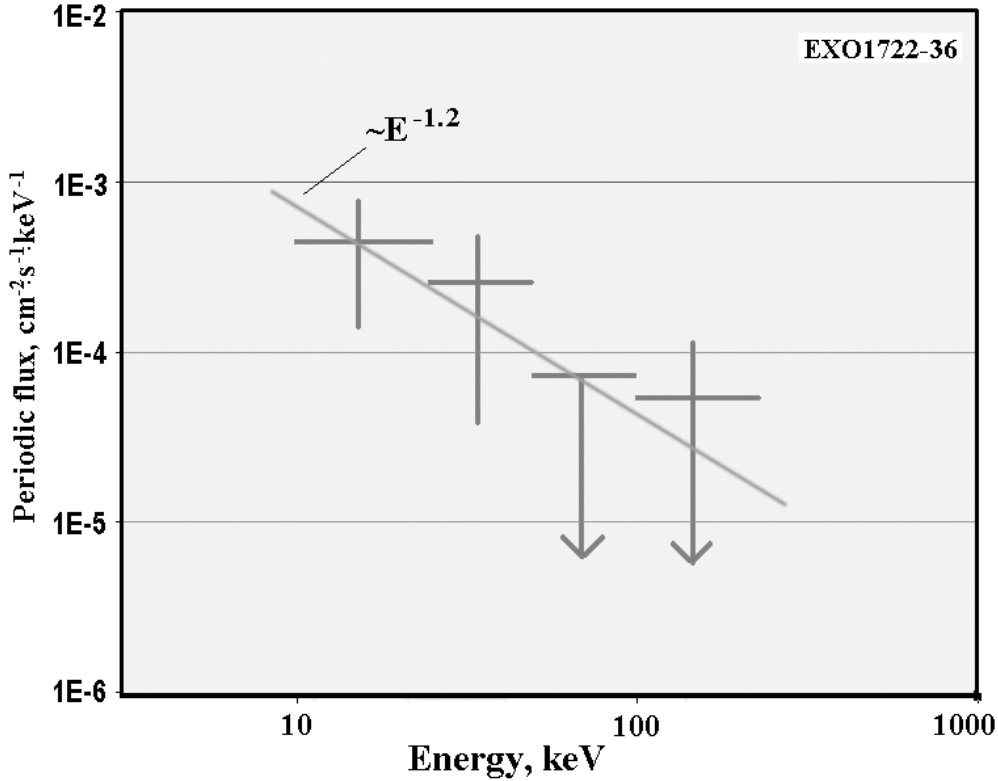}}
      \caption{ Energy spectra of GS1722-36 pulsed radiation obtained in 
       Prognoz 9 experiment for TJD5668.40-5675.31 interval of observation. 
       The line corresponds to the best-fit power law approximation of the 
       spectrum with the slope $\sim 1.2$.}
         \label{Fig8}
   \end{figure}

There is no any information about GS1722-36 orbital phases in the epoch of Prognoz 9 
observations, thus we could not make evident conclusions about the stage of the source, when 411.7 
s periodic process was observed. However, taking into account the probable distance to the source – 
about 10 kps, the luminosity in the pulsed component may be obtained at the level $\sim 5*10^{36}$ 
erg/s, which is typical for the ordinary (non-giant) outburst. This value is similar to one obtained 
in Ginga experiment \citep{Taw89}.

	In the "Mir" GRIF experiment the upper limits on the GS1722-36 pulsation flux in 25-50 
keV were obtained (see table 1). The source was observed during four time intervals. As it could be 
seen from the Table 1 the pulsation flux upper limits in all intervals are of the same order than the 
full flux according to the Integral data \citep{Lut04a}. Due to the rather short orbital period ($\sim 9.7 d$) 
\citep{Lut04b} it is difficult to estimate the orbital phases, which corresponded to those intervals, which time was 
longer than orbital period or more than one half of it value, as in the case of the last interval 
(10410.0 – 10415.0 TJD). Thus, the presented in Table 1 upper limits characterize the pulsation 
intensity averaged over the orbit. Their quite ordinary values may indicate on that there were no any 
giant outburst during those time intervals.

\section{Discussion}

As it could be seen from the Tables 1 and 2, the values of pulsation flux upper limits 
obtained in the Prognoz 9 and GRIF "Mir" experiments are in the range of typical variability of 
correspondent transient pulsars. Taking into account flux variability, these data can be used for the 
estimation of the transient pulsar activity for different epochs of observations. Besides the 
considered above A0535+262 and GS1722-36 the data about seven transient pulsars (4U1145-619, 
A1118-615, EXO2030+37, Sct X-1, SAX J2103.5+4545, IGR 16320-4751, IGR 16465-4507) are 
also presented. By the observational conditions only one of them (Sct X-1) was observed in both 
experiments, others were visible only in the GRIF X-ray spectrometer FOV. 
For the time of observation of three pulsars (4U1145-619, A1118-615, EXO2030+37) in the 
GRIF experiment the BATSE CGRO data as well as the information about those binary systems 
orbital parameters were also accessible \citep{Bil97}. It allows to compare the pulsation 
flux values in the near energy ranges and to connect the time of the GRIF observations with  
corresponding orbital phase, if this time was not so long to surpass the half of the orbital period. As 
for the Sct X-1, the only fragmentary observations are known (), there were no any other 
experiments, which could be intersected in time with Prognoz 9 and GRIF observations. Thus, it 
makes difficult to realize any comparison on the pulsation flux limit and orbital phase for this 
source. The last three pulsars (SAX J2103.5+4545, IGR 16320-4751, IGR 16465-4507) are the 
newly discovered objects \citep{Hul98,Mol04} and there were no observations near in time to the GRIF experiment. 
As the result, for those pulsars we have no information about the orbital phases corresponding to the 
times of GRIF observations. However for two of them (IGR 16320-4751, IGR 16465-4507) the 
Integral data are already known \citep{Lut04a,Mol04}, which allow to estimate their fluxes.
The upper limits of pulsation flux obtained in the GRIF experiment for IGR 16320-4751 and 
IGR 16465-4507 transient pulsars do not exceed their full fluxes according to the Integral data. The 
presented values as well as the pulsation flux upper limits for SAX J2103.5+4545 and Sct X-1 
indicate, probably, on those pulsars were in the quiescent state or in the finale stage of the first type 
outburst, because the main stage of the ordinary outbursts and second type (giant) outbursts 
are usually characterized by sufficiently higher fluxes.

	Those pulsars, for which it was possible to 
determine orbital phases, were evidently observed between outbursts (orbital phase 
$>$0.3) or in the outburst decreasing state (orbital phase 0.14-0.5). If to assume the probable distance 
to these sources about few kps, we may conclude that typical flux upper limits from Tables 1 and 2 
about $(3-5)\cdot 10^{-10} erg/cm^2\cdot s$ correspond to the limits on the luminosity in the source 
$\sim 10^{35}$ erg/s, which are indeed of the order of transient pulsar luminosity between outbursts. 
	Thus, the given overview of the estimations of the different pulsars activity from the long-
term observations made in the Prognoz 9 and GRIF space X-ray experiments confirm the accepted 
point on the transient pulsar activity, according to which the regular intensive raises of fluxes (first 
type outbursts) associated with periastron passage (orbital phase 0 - 0.3). The results of Prognoz 9 and 
GRIF experiments indicate the absence of the giant (second type) outbursts in correspondent epoch of observation.

	The mean phase profiles of pulsations of the detected transient pulsars A0535+26 and GS1722-36 
demonstrate the general similarity of their phase curves with ones obtained in other 
experiments. However some significant difference in detailes is present. It is still unclear if this difference is
caused by qualitative changes of accretion region geometry in transient pulsar binary system from one normal 
outburst to another or the variations in total amount of accreting matter can explain all the observational data.
It is important to understand if there is any systematic evolution of the phase profiles shape from one epoch 
to another. Additionally the energy dependency of the shape of phase profile as well as its dependency on the 
total or pulsed x-ray intensity in the different stages of normal outbursts are of the great interest 
and may become the theme of further exploration.

\bibliographystyle{aa}

\end{document}